\documentclass[a4paper]{PoS}

\title{(Dimensional) twisted reduction in large N gauge theories}
  
\ShortTitle{(Dimensional) twisted reduction in large N gauge theories}

\author{Liam Keegan\\
        PH-TH, CERN, CH-1211 Geneva 23, Switzerland\\
        E-mail: \email{liam.keegan@cern.ch}}

\author{\speaker{Alberto Ramos}\\
        PH-TH, CERN, CH-1211 Geneva 23, Switzerland\\
        E-mail: \email{alberto.ramos@cern.ch}}

\abstract{We show that the spontaneous breaking of center symmetry can
be avoided on a $L^2\times 1^2$ lattice with the appropriate choice of
twisted boundary conditions. In order for this to work it is crucial
that the twisted boundary conditions are chosen in the reduced
plane. This suggests that the choice of twist tensor can influence the
directions in which color and space degrees of freedom become
indistinguishable. We also present some preliminary quantitative data
comparing the value of the plaquette for different forms of reduction.
\begin{flushright}
CERN-PH-TH-2015-254
\end{flushright}
}

\FullConference{The 33rd International Symposium on Lattice Field Theory\\
                 14 -18 July  2015\\
                 Kobe International Conference Center, Kobe, Japan}

\usepackage{multirow}
\usepackage{graphicx}
\usepackage{float}
\usepackage{amsmath}
\usepackage{multicol}
\usepackage[margin=4pt]{subcaption}
\begin{document}

\section{Introduction}

$SU(N)$ Yang-Mills (YM) theories become volume independent in the
large $N$ limit~\cite{Eguchi:1982nm}. In the context of lattice
gauge theories it allows, in principle, to use a one-site lattice
to obtain results in the large $N$
limit and infinite volume. 
If implemented in the most naive way, this
proposal was shown to fail~\cite{Bhanot:1982sh}, the reason being that
the proof of volume independence relies on center symmetry, while for
very small volumes (below the confinement scale), center symmetry is
spontaneously broken. The literature contains several proposals to
save the idea of volume 
independence~\cite{Bhanot:1982sh,GonzalezArroyo:1982ub,GonzalezArroyo:1982hz,
GonzalezArroyo:2010ss,Kiskis:2003rd,Kovtun:2007py,Basar:2013sza,Unsal:2008ch,Azeyanagi:2010ne}. 
In this work we will focus on the twisted reduction
approach~\cite{GonzalezArroyo:1982ub,GonzalezArroyo:1982hz,Teper:2006sp,Azeyanagi:2007su,GonzalezArroyo:2010ss}. The   
key idea is that the behavior of a field theory at small volumes
depends crucially on the choice of boundary conditions, while the
proof of volume reduction does not. In particular
it has been shown recently that a judicious choice of twisted boundary
conditions can avoid the spontaneous breaking of center symmetry. The
choice of twisted boundary conditions are encoded in some phases
$z_{\mu\nu} = \exp(2\pi\imath n_{\mu\nu}/N)$, where $n_{\mu\nu}$ is a
tensor of integers modulo $N$. 

The work~\cite{GonzalezArroyo:1982hz} correctly argued that any
choice of twist 
tensor $n_{\mu\nu}$ could do the job as long as it prevents the
breaking of center symmetry. The study of this model at weak coupling
suggests the particular choice of taking $N$ a perfect square $N=\hat
L^2$ and the symmetric twist tensor $n_{\mu\nu} = k\hat L$
where $k$ and $\hat L$ are co-prime. 
The reason for the notation $\hat
L$ is that the Feynman rules and propagators in the one-site model
are identical to the Feynman rules on a lattice with $\hat L$ points
in each dimension, except for some momentum-dependent phases that
cancel in all non-planar diagrams. In this way the $\mathcal O(N^2)$
color degrees of 
freedom of the gluons are indistinguishable from the usual
space degrees of freedom on a $\hat L^4$ lattice in the large $N$ limit. 

In this work we will explore the possibility of a non-symmetric
twist. We will chose to twist only one plane (say the $x_1-x_2$
plane), while our gauge fields will be strictly periodic in the other
two directions $x_0,x_3$. We will argue that in this case the $\mathcal
O(N^2)$ color degrees of freedom transform in space-time
degrees of freedom only into the twisted plane $x_1-x_2$. This will
allow us to reproduce $SU(\infty)$ results in the infinite volume
limit by exploring the large $N$ limit of a theory with two short
directions ($x_1,x_2$), and two infinite directions ($x_0,x_3$). We
will call this the 2d TEK model. We will show that the pattern of 
spontaneous symmetry breaking will be consistent with this
expectation: the strictly periodic directions $x_0,x_3$ must be kept
``long'' in order to avoid symmetry breaking\footnote{In this work we
  will not explore the interesting possibility of scaling the bare
  parameters in order to obtain a continuum non-commutative field
  theory. See~\cite{Bietenholz:2006cz} for this approach.}.

\section{Choice of twisted boundary conditions and simulation details}

A comprehensive review of the twisted boundary conditions is beyond
the scope of this work. We will stick to the basics and refer the
reader to the existing literature~\cite{ga:torus,Perez:2014sqa}. 

We are going to work in a four dimensional torus $\mathcal T^4$ of
sides $L_0\times L_1\times L_2\times L_3$. Twisted boundary conditions
are imposed by requiring
\begin{equation}
  A_\mu(x+L_\nu\hat \nu) = \Omega_\nu(x)A_\mu(x)\Omega^+_\nu(x) + 
  \Omega_\nu(x)\partial_\mu \Omega_\nu^+(x)\,,
\end{equation}
where matrices $\Omega_{\mu}(x)$ are called twist matrices, and they
have to obey the consistency relation
\begin{equation}
  \Omega_\mu(x+L_\nu\hat\nu)\Omega_\nu(x) = z_{\mu\nu}
  \Omega_\nu(x+L_\mu\hat\mu)\Omega_\mu(x) 
\end{equation}
with $z_{\mu\nu}$ being elements of the center of $SU(N)$. They are 
gauge invariant, and therefore encode the physical part of the
twisted boundary conditions. We are going to use a particular setup of
this general scheme: we twist the plane $x_1,x_2$, while
the gauge potential will be periodic in the directions $x_0,x_3$. This
amounts to the choice
\begin{equation}
  z_{\mu\nu} = z^*_{\nu\mu} = \left\{
    \begin{array}{ll}
      \exp(2\pi\imath k/N) & \mu=1\, {\rm and }\, \nu=2 \\
      1 & {\rm otherwise} \\
    \end{array}
  \right.\,.
\end{equation}
Finally if we choose the integer $k$ to be co-prime with $N$, it is
guaranteed that there is a unique field configuration that minimizes
the action modulo gauge transformations (i.e. the only zero-modes of
the action are the gauge degrees of freedom). This is crucial for our
purposes, since constant configurations that are minima of the
action (``torons'') are the source of spontaneous center symmetry
breaking.  

It can be proved~\cite{ga:torus} that any gauge connection compatible
with these particular boundary conditions can be written as
\begin{equation}
  \label{eq:connection}
  A_\mu^a(x)T^a = \frac{1}{\prod_\mu L_\mu} \sum_{\tilde p\ne 0}\tilde
  A_\mu(x,\tilde p)e^{\imath  
    \tilde px}\hat\Gamma(\tilde p),
\end{equation}
where $\tilde A_\mu(x,\tilde p)$ are complex functions (\emph{not}
matrices) periodic in $x$, and $\hat\Gamma(\tilde p)$ are some well
defined matrices (see~\cite{ga:torus} for more details). Finally the
color-momentum $\tilde p_\mu$ is defined as 
\begin{equation}
  \tilde p_\mu = \frac{2\pi \tilde n_\mu}{NL_\mu}\,, \quad 
  \tilde n_\mu = \left\{
    \begin{array}{ll}
      0 & \mu = 0,3 \\
      0,\dots,N-1 & \mu = 1,2 \\
    \end{array}
  \right.\,.
\end{equation}

Note that the function $\tilde A_\mu(x,\tilde p)$, being periodic in
the torus, has naturally momentum modes quantized in units of
$2\pi/L_\mu$, but 
Eq.~(\ref{eq:connection}) suggests that actually in the directions
$x_1,x_2$ the unit of quantization of the momentum is $2\pi/NL_\mu$
(i.e. the gauge field lives in an apparently larger lattice of size
$NL$ in the $x_1,x_2$ directions). 

We still have to choose $k$. The experience with the TEK model and
symmetric twist suggests that keeping $k=1$ would result
in the spontaneous breaking of center symmetry~\cite{Teper:2006sp,Azeyanagi:2007su}. We
will see that the same holds here, but the advice
of~\cite{GonzalezArroyo:2010ss} (i.e. $k/N>1/9$) will also avoid the
spontaneous breaking of center symmetry in our case. Moreover, based
on the experience with the symmetric twist it seems reasonable
to keep $\bar k/N$ as constant as possible, where $\bar k$ is defined
by the condition $k\bar k = 1\mod N$.
The parameters
that we will use in our runs are specified in
Table~\ref{tab:parms}. All simulations are done using standard
over-relaxation techniques for $SU(N)$ after introducing auxiliary
variables to linearize the
action~\cite{Vairinhos:2010ha,Perez:2015ssa}. 

\begin{table}
  \centering
  \begin{tabular}{l|llll}
    \hline
    $N$ & 24 & 36 & 40 & 56 \\
    $k$ & 7  & 13 & 11 & 23 \\
    $\bar k/N$ & 0.291666\dots&0.305555\dots&0.275&0.303571\dots  \\
    \hline
 \end{tabular}
  \caption{Parameters of the runs}
  \label{tab:parms}
\end{table}
\section{Pattern of spontaneous symmetry breaking}

\begin{figure}[ht]
   \centering
   \includegraphics[width=\textwidth,height=0.3\textheight]{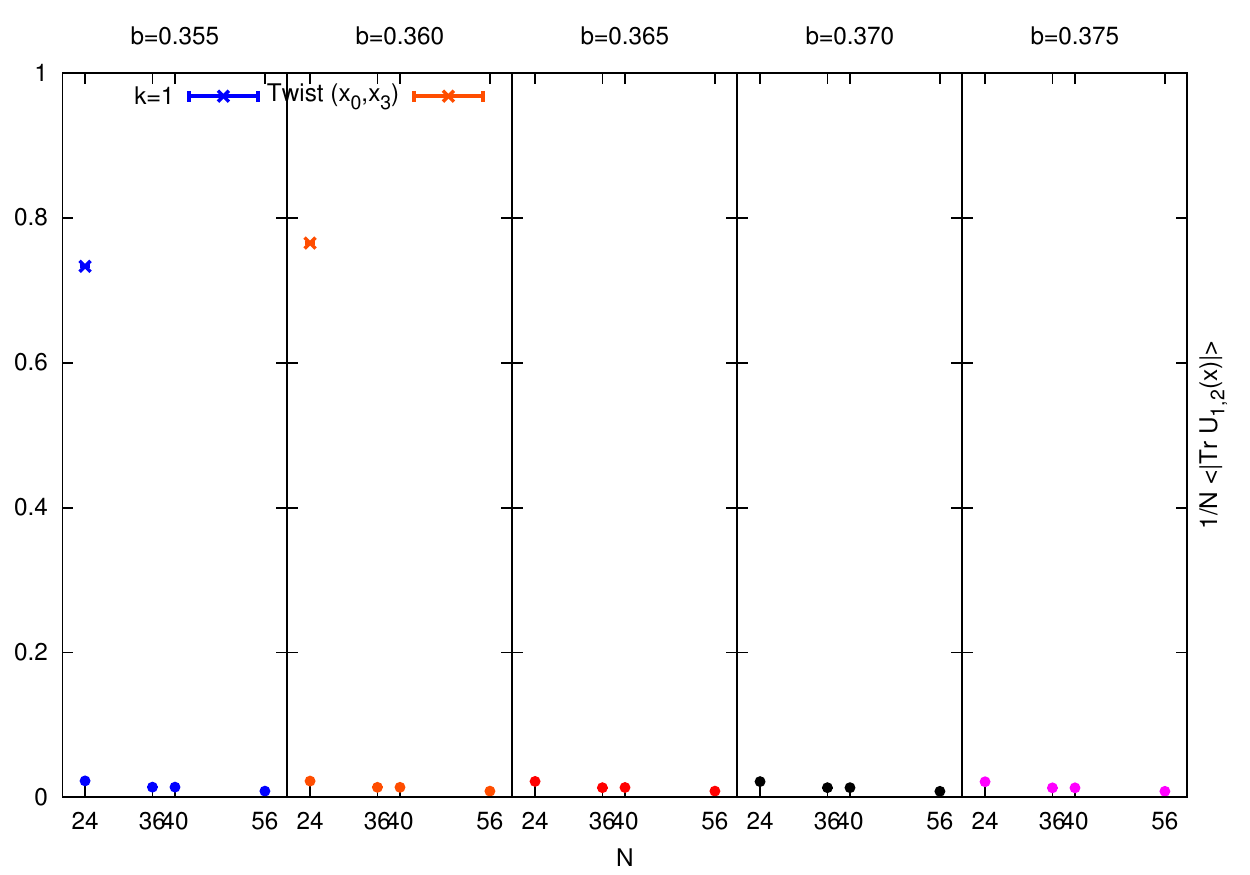}
   \caption{$\frac{1}{N}\langle|{\rm Tr}U_{1,2}(x)|\rangle$ for
     different values of $b$ and $N$ in a simulation of $SU(N)$ on a
     $N\times 1\times 1\times N$ lattice. Circles represent results
     obtained with our run parameters, and
     crosses some test runs: one with $k=1$ and another using twisted
     boundary conditions in the $(x_0,x_3)$ plane.}
   \label{fig:zn}
 \end{figure}

Pushing the idea of volume independence to the limit, we should be
able to obtain continuum results of $SU(\infty)$ in $\mathbb
R^4$ by computing any observable on a $L_0/a\times 1\times
1\times L_3/a$ lattice. 
The $N\rightarrow \infty$ limit has to be taken at
fixed bare coupling, and only later the continuum limit ($b\rightarrow
\infty$ keeping a line of constant physics). It is easy to understand
why center symmetry must be preserved: on an infinite lattice, open
paths are protected from getting an expectation value by gauge
invariance. On  
the other hand on our reduced lattice, the quantities ${\rm Tr}U_i(x)$
for $i=1,2$ (i.e. the reduced directions in which the lattice has only
one point) are gauge invariant, although not center invariant. The
preservation of center symmetry in the reduced model can therefore be
related with the preservation of gauge invariance on the infinite
lattice. This naturally suggests to use as order parameters to study
the breaking of center symmetry the quantities~\cite{Teper:2006sp}
\begin{equation}
  P_n = \frac{1}{N}\langle |{\rm Tr}U_i^n(x)| \rangle; \quad
  Q_{n,m} = \frac{1}{N}\langle |{\rm Tr}U_i^n(x)U_j^m(x)| \rangle\,.
\end{equation}
These quantities have to be zero\footnote{More precisely, they have to
go to zero as $N\rightarrow\infty$.} for all $0<n,m<N$. The most
stringent constraint comes in fact from $P_1$, and this is the case
that we will study in more detail. Figure~\ref{fig:zn} shows the
comparison of the cases in which center symmetry is spontaneously
broken, and cases in which this is not the case. The points are
results of the simulations with the parameters of
table~\ref{tab:parms} and the choice $L_{0,3}/a=N$. As the reader can
observe, this choice of $k$ 
(following~\cite{GonzalezArroyo:2010ss}) avoids the breaking of center
symmetry. On the other hand the same plot shows that
the value of $k$ plays a crucial role in avoiding the breaking of
center symmetry: simulations with $k=1$ shows this breaking. Also the
choice of which directions are large and which ones are short cannot
be arbitrary: changing the twist from the plane $x_1-x_2$ to the plane
$x_0-x_3$ but keeping the same geometry (i.e. $L_{1,2}/a=1$ and
$L_{0,3}/a=N$) shows again signs of spontaneous symmetry
breaking. 

\section{Comparison with the TEK model with symmetric twist}

Following the above discussion, results of the twisted reduced model
with different choices of twist should agree in the large $N$
limit. In this section we will compare the usual one-site TEK model
(which we will call 0d TEK) with symmetric twist with the choice of
twist described above that we called 2d TEK model.  

\begin{table}
  \centering
  \begin{tabular}{l|llll}
    \hline
    $N$& $b=0.355$&$b=0.360$&$b=0.365$&$b=0.370$ \\
    \hline
    24& 0.546681(11) &0.5592981(82) &0.5702507(67)&0.5801170(77)\\
    36& 0.5458441(63)&0.5585279(46) &0.5695444(41)&0.5794575(39)\\
    40& 0.5458078(47)&0.5584742(45) &0.5694683(31)&0.5793679(31)\\
    56& 0.5455784(56)&0.5582325(52) &0.5692438(46)&0.5791585(47)\\
    \hline
    $\infty$ (d=2) &0.545326(40)&0.557988(35)&0.569018(17)&0.5789434(64)\\
    $\infty$ (d=4$^*$)&0.545417(63)&0.558012(12)&0.569021(41)&0.578978(17)\\
    $\infty$ (d=0$^*$)&0.545336(11)&0.558019(11)&0.569018(4)&0.578959(5)\\
    \hline
  \end{tabular}
  \caption{Values of the average plaquette. Values with $N=\infty$
    corresponds to extrapolated values. Values marked with an $^*$ are
  taken from~\cite{Gonzalez-Arroyo:2014dua}. See text for more details.}
  \label{tab:plaq}
\end{table}

\begin{figure}[h]
  \centering
  \begin{subfigure}[t]{0.48\textwidth}
    \centerline{\includegraphics[width=\textwidth]{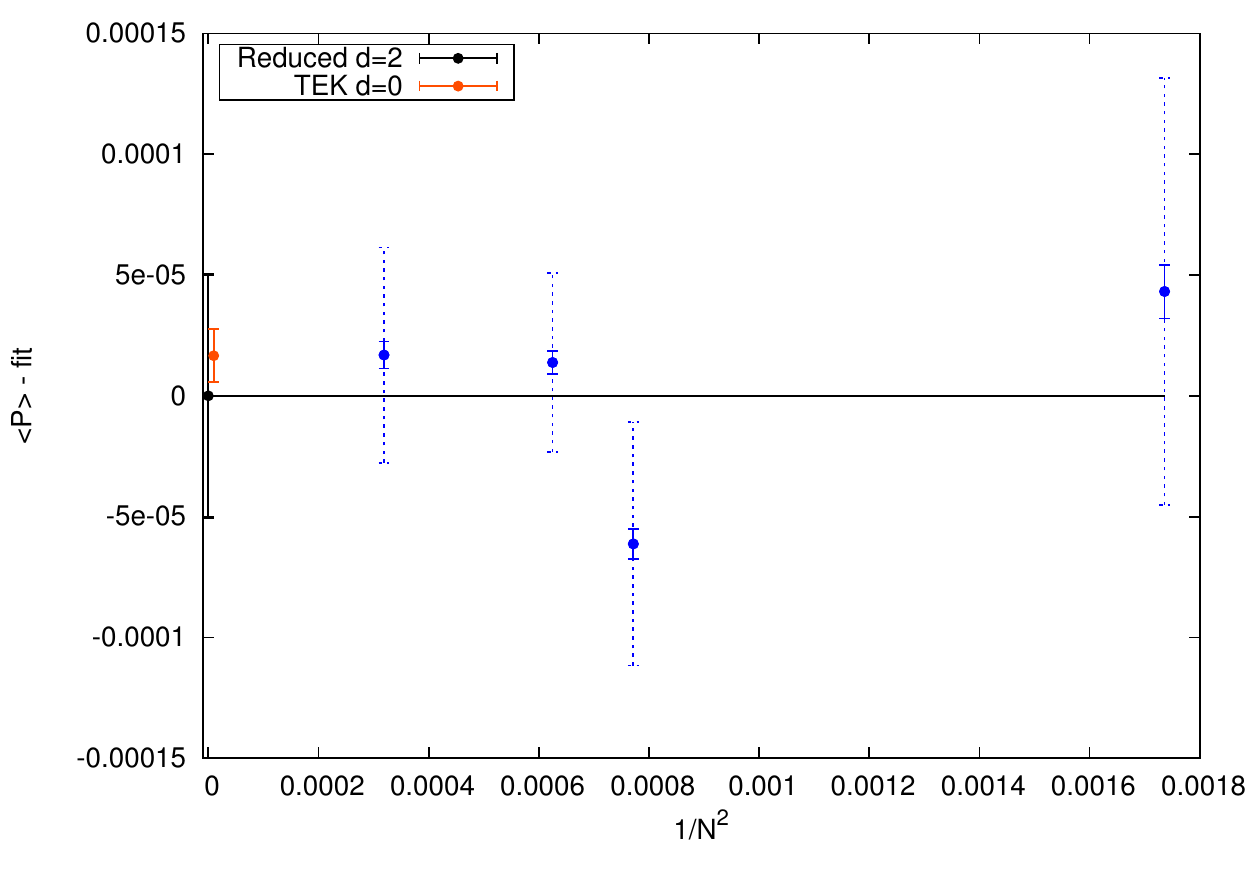}}
    \caption{$b=0.355$}
    \label{fig:oscb0355}
  \end{subfigure}
  \begin{subfigure}[t]{0.48\textwidth}
    \centerline{\includegraphics[width=\textwidth]{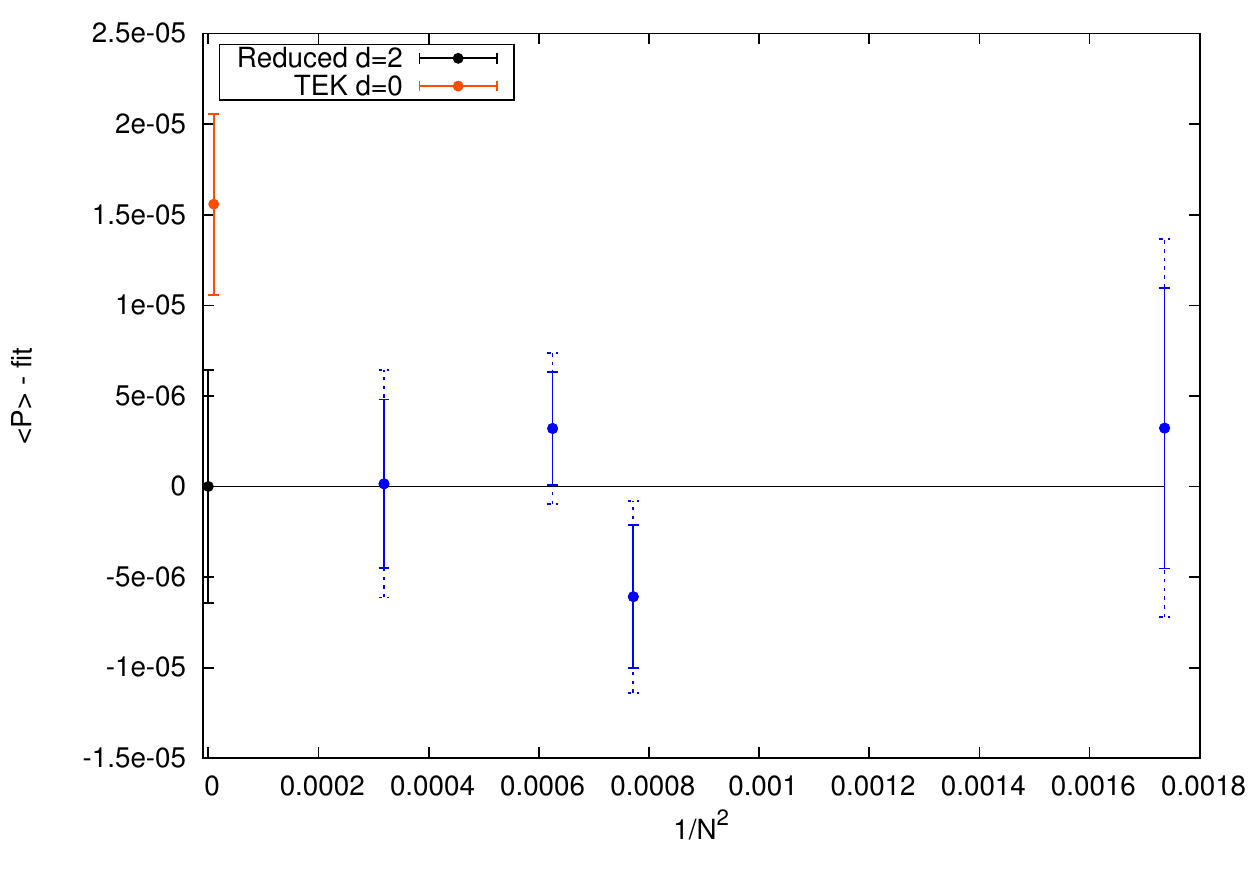}}
    \caption{$b=0.370$}
    \label{fig:oscb0370}
  \end{subfigure}
  \caption{Oscillations around the $1/N^2$ scaling for $b=0.355$ and
    $b=0.370$. Solid error bars are the statistical errors. The
    difference between the dotted and the solid error bars are an
    estimation of the systematic uncertainty due to the variation of
    $k, N$ and $L/a$. Note that this difference is smaller closer to
    the continuum.} 
  \label{fig:scaling}
\end{figure}

\begin{figure}[h]
  \centering
  \begin{subfigure}[t]{0.48\textwidth}
    \includegraphics[width=\textwidth]{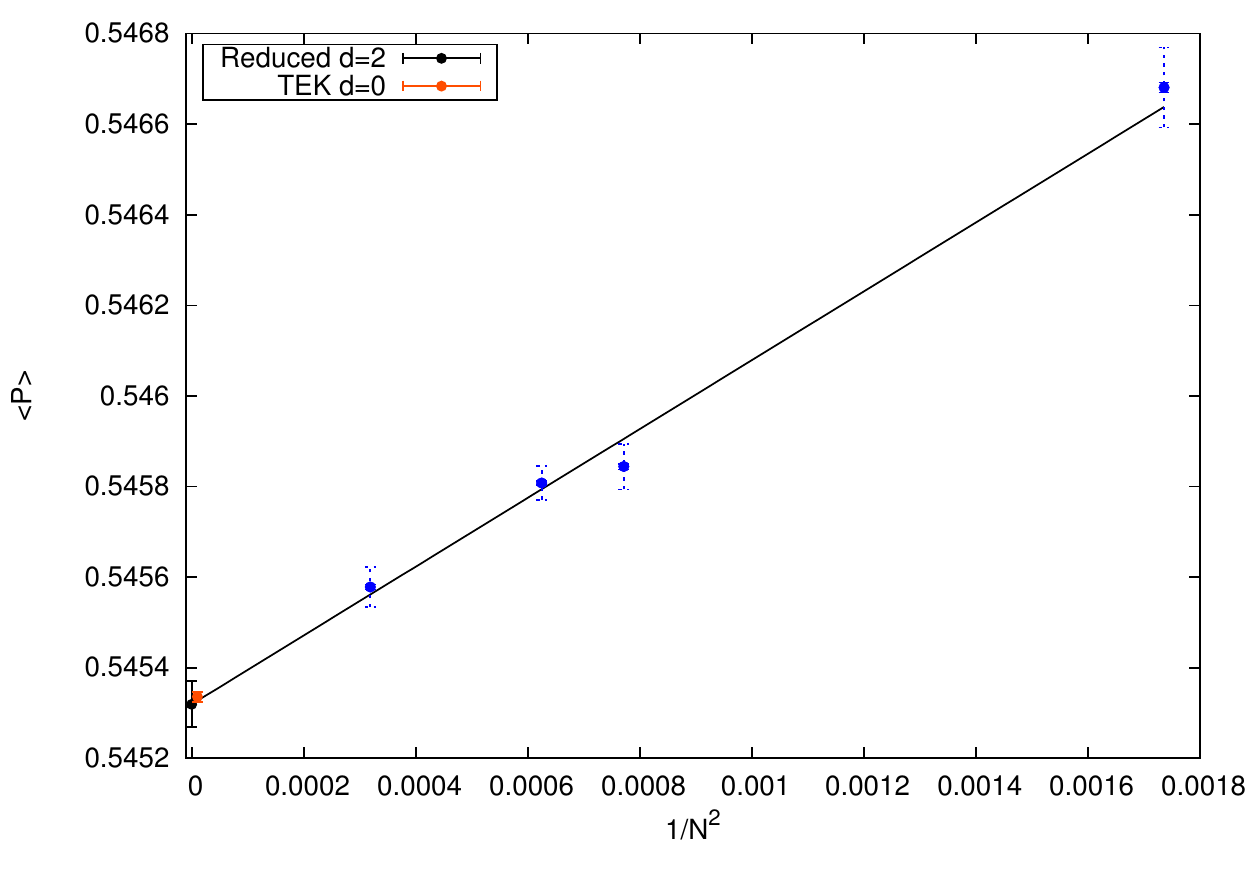}
    \caption{$N\rightarrow\infty$ extrapolation (data of
      Table~\ref{tab:plaq}). The
      statistical 
      error in the data is approximately of the size of the
      points. The dotted error bars are estimated with the fit quality
      (see text for more details).}
    \label{fig:plaq}
  \end{subfigure}
  \begin{subfigure}[t]{0.48\textwidth}
    \includegraphics[width=\textwidth]{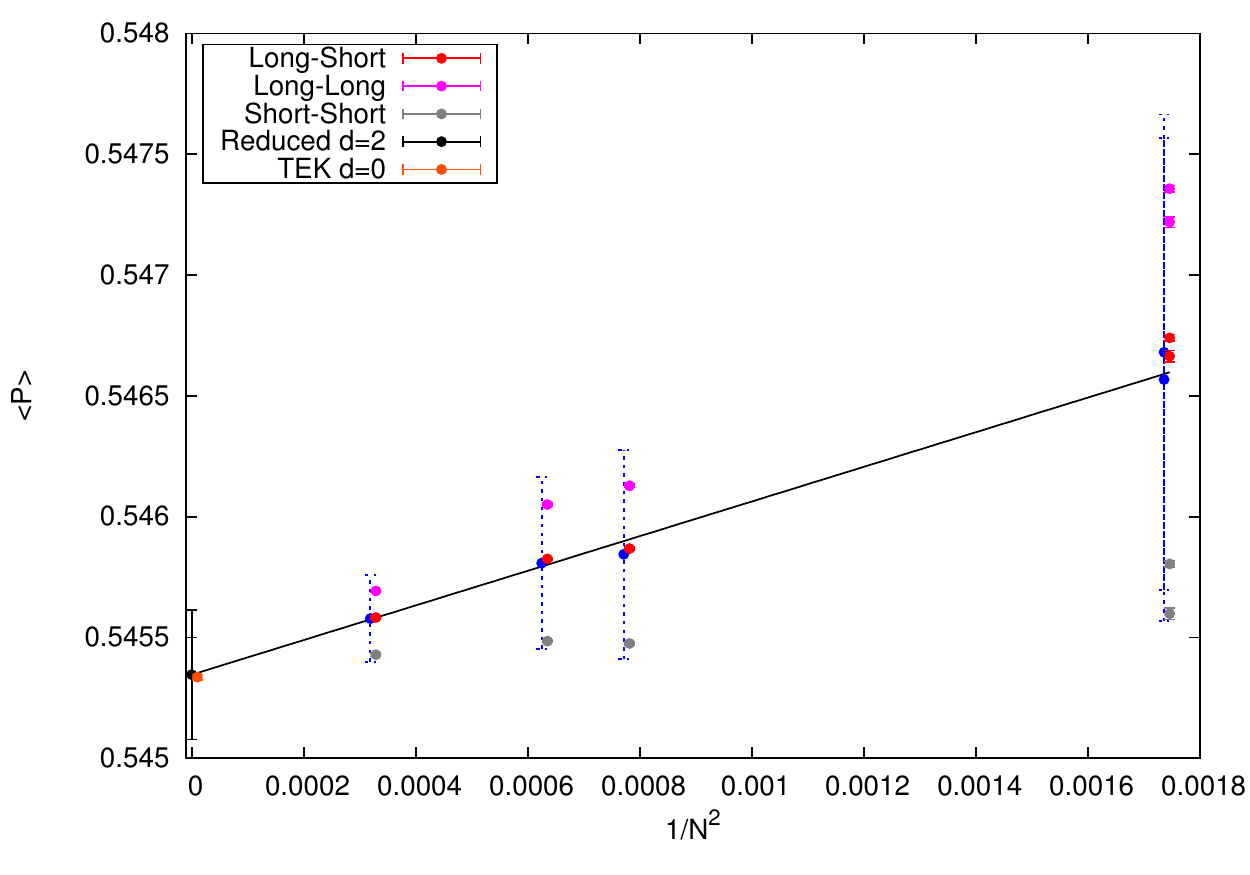}
    \caption{Average plaquette values for each plane. The spread
      between planes can be used to estimate how far observables at 
    finite $N$ are from the $N=\infty$ limit.}
    \label{fig:pl}
  \end{subfigure}
  \caption{Average plaquette values for $b=0.355$}
  \label{fig:b0p355}
\end{figure}

For this preliminary study we will choose as observable to perform the
comparison the plaquette. Figure~\ref{fig:plaq} shows the
extrapolation of our data for the case of $b=0.355$. 
In general the finite $N$ corrections to any quantity will 
depend not only on $N$, but also on the choice of parameters $L/a$
and $k$. How to take into account the variations of $k,N$ and $L/a$ in
the $N\rightarrow\infty$ extrapolation requires a deeper understanding
of the general structure of the corrections. For the case of the $0d$
TEK model, the corrections are parametrized by the quantity $\bar
k/\sqrt{N}$~\cite{Perez:2014sqa}. This suggests to keep constant the
ratio $\bar k/N$ for the case of the $2d$ twisted reduction. We have
chosen our parameters 
(Table~\ref{tab:parms}) according to this criteria, but still we can
observe some corrections to a perfect $1/N^2$ scaling. For example, 
Figures~\ref{fig:scaling} show the difference between our data points
and a linear $1/N^2$ fit. The solid lines are the statistical errors
of the data points. The dotted lines represent an error bar the
includes an estimate of the
uncertainty due to our choices of $k,N$ and $L/a$. This estimate is
computed by using the $\chi^2/{\rm dof}$ to weight the errors of the
points in such a way that the fit becomes good (i.e. we multiply all
error bars by a factor such that the $\chi^2/{\rm dof}$ becomes 1). An
important observation is that this effect is less important closer to
the continuum. 

Table~\ref{tab:plaq} summarizes the values of the plaquette. The
values at $N=\infty$ corresponds to extrapolations using
different setups. The values quoted as $d=4$ correspond to the naive
extrapolation to $N\rightarrow\infty$ on a 4d lattice with linear size
$L/a=16$. The values quoted as $d=0$ corresponds to results of the
TEK model with symmetric twist. Both these values are taken from the
existing literature~\cite{Gonzalez-Arroyo:2014dua}. Finally the values
at finite $N$ and the extrapolation quoted $d=2$ corresponds to our
setup with two large directions and two reduced ones. As the reader
can see the quantitative agreement of the three scenarios is very
good. 

Fig.~\ref{fig:pl} shows the average plaquette values for each
plane. In our simulations we have 3 types of planes, one made by
short directions ($x_1-x_2$), one made by long directions ($x_0-x_3$),
and four planes that mix the long and the short directions. The final
plaquette values quoted in Table~\ref{tab:plaq} are the average over
all the planes, but since in the infinite volume and infinite $N$
limit all planes are equivalent, the spread between these values can
be used to estimate how far a certain simulation is from this
limit. Figure~\ref{fig:pl} shows that indeed this is the case: 
the spread between 
different planes gets reduced when $N$ increases. This is a sign that
the $O(4)$ invariance of the infinite volume and infinite $N$ limit
is restored despite the fact that our simulations all have
$L_1/a=L_2/a=1$. 
\vspace*{-1cm}

\section{Conclusions}

In the context of twisted
reduction~\cite{GonzalezArroyo:1982ub,GonzalezArroyo:1982hz} we have
studied the 
possibility of using a non-symmetric twist. We have shown that in
order to avoid the spontaneous breaking of center symmetry (a
necessary condition for reduction to hold), we have to keep the two
strictly periodic directions large, while the directions on the
twisted plane can be made arbitrarily small with a proper choice of
twist tensor. In this sense the choice of twist tensor allows to make
color degrees of freedom indistinguishable from space degrees of
freedom in certain directions. 

Beyond the theoretical interest in this form of twisted reduction,
this choice of twist tensor might have some advantages. Among them one
of the long directions can be naturally interpreted as the time
direction, allowing 
to compute masses by looking at the large euclidean time
behavior of correlators (see also~\cite{Gonzalez-Arroyo:2015bya}). 
The reader should also note that the cost of matrix 
multiplications grows like $\mathcal O(N^3)$, while the
computations on a lattice grows with the volume (therefore like
$\mathcal (L/a)^2$ for the non-reduced directions), making this choice of
twisted reduction cheaper from a numerical point of view. Also the
lattice structure provides a natural way to perform distributed
computations. The main
drawback is that, having to keep the space degrees of freedom in the
simulations one can reach only smaller values of $N$, and therefore an 
extrapolation to 
$N\rightarrow\infty$ is still required. A deeper understanding of how
the choices of $k,N$ and $L/a$ affect the $N\rightarrow\infty$ limit
would be desirable.

\section{Acknowledgments}

The authors want to show their gratitude to M. Garcia P\'erez and
A. Gonz\'alez-Arroyo for the many illuminating discussions and for
sharing some of their results before publication. This project, being
also a small programming experiment, has profited from the help and
work of Alessandro Fanfarillo and the other developers of the
\texttt{opencoarray} library~\cite{opencoarray:2014}. The help and
patience of the staff of the University of Cantabria at IFCA, specially
I. Campos, was crucial for this project. Computations were
performed in the ALTAMIRA HPC cluster at the IFCA.  

\makeatletter
\renewenvironment{thebibliography}[1]
     {\begin{multicols}{2}[\section*{\refname}]%
      \@mkboth{\MakeUppercase\refname}{\MakeUppercase\refname}%
      \list{\@biblabel{\@arabic\c@enumiv}}%
           {\settowidth\labelwidth{\@biblabel{#1}}%
            \leftmargin\labelwidth
            \advance\leftmargin\labelsep
            \@openbib@code
            \usecounter{enumiv}%
            \let\p@enumiv\@empty
            \renewcommand\theenumiv{\@arabic\c@enumiv}}%
      \sloppy
      \clubpenalty4000
      \@clubpenalty \clubpenalty
      \widowpenalty4000%
      \sfcode`\.\@m}
     {\def\@noitemerr
       {\@latex@warning{Empty `thebibliography' environment}}%
      \endlist\end{multicols}}
\makeatother
\let\oldbibliography\thebibliography
\renewcommand{\thebibliography}[1]{\oldbibliography{#1}
\setlength{\itemsep}{-2pt}} 

\bibliography{/home/alberto/docs/bib/math,/home/alberto/docs/bib/campos,/home/alberto/docs/bib/fisica,/home/alberto/docs/bib/computing}

\providecommand{\href}[2]{#2}\begingroup\raggedright\begin{thebibliography}{10}

\bibitem{Eguchi:1982nm}
T.~Eguchi and H.~Kawai, 
{\em Phys.Rev.Lett.} {\bf 48} (1982) 1063.

\bibitem{Bhanot:1982sh}
G.~Bhanot at al.
{\em Phys.Lett.} {\bf B113} (1982) 47.

\bibitem{GonzalezArroyo:1982ub}
A.~Gonzalez-Arroyo and M.~Okawa, 
{\em Phys.Lett.} {\bf B120} (1983) 174.

\bibitem{GonzalezArroyo:1982hz}
A.~Gonzalez-Arroyo and M.~Okawa, 
{\em Phys.Rev.} {\bf D27}
  (1983) 2397.

\bibitem{GonzalezArroyo:2010ss}
A.~Gonzalez-Arroyo and M.~Okawa, 
{\em JHEP} {\bf 1007} (2010) 043,
  [\href{http://xxx.lanl.gov/abs/1005.1981}{{\tt arXiv:1005.1981}}].

\bibitem{Bietenholz:2006cz} 
  W.~Bietenholz et al.
  JHEP {\bf 0610}, 042 (2006)
  [hep-th/0608072].

\bibitem{Kiskis:2003rd}
J.~Kiskis et al.
 {\em Phys.Lett.} {\bf B574} (2003) 65--74,
  [\href{http://xxx.lanl.gov/abs/hep-lat/0308033}{{\tt hep-lat/0308033}}].

\bibitem{Kovtun:2007py}
P.~Kovtun et al.
{\em JHEP} {\bf 0706} (2007) 019,
  [\href{http://xxx.lanl.gov/abs/hep-th/0702021}{{\tt hep-th/0702021}}].

\bibitem{Basar:2013sza}
G.~Basar et al.
{\em
  Phys.Rev.Lett.} {\bf 111} (2013), no.~12 121601,
  [\href{http://xxx.lanl.gov/abs/1306.2960}{{\tt arXiv:1306.2960}}].

\bibitem{Unsal:2008ch}
M.~Unsal and L.~G. Yaffe, 
{\em Phys.Rev.} {\bf D78}
  (2008) 065035, [\href{http://xxx.lanl.gov/abs/0803.0344}{{\tt
  arXiv:0803.0344}}].

\bibitem{Azeyanagi:2010ne}
T.~Azeyanagi et al.
{\em Phys. Rev.} {\bf
  D82} (2010) 125013, [\href{http://xxx.lanl.gov/abs/1006.0717}{{\tt
  arXiv:1006.0717}}].

\bibitem{Teper:2006sp}
M.~Teper and H.~Vairinhos, 
{\em Phys. Lett.} {\bf B652} (2007) 359--369,
  [\href{http://xxx.lanl.gov/abs/hep-th/0612097}{{\tt hep-th/0612097}}].

\bibitem{Azeyanagi:2007su} 
  T.~Azeyanagi et. al.
  JHEP {\bf 0801}, 025 (2008)
  [arXiv:0711.1925 [hep-lat]].


\bibitem{ga:torus}
A.~Gonz{\'a}lez-Arroyo, 
{\em {World Scientific.}}
  (1998) {Singapore}, [\href{http://xxx.lanl.gov/abs/hep-th/9807108}{{\tt
  hep-th/9807108}}].

\bibitem{Perez:2014sqa}
M.~G. Perez et al.
  \href{http://xxx.lanl.gov/abs/1406.5655}{{\tt arXiv:1406.5655}}.

\bibitem{Vairinhos:2010ha}
H.~Vairinhos, 
  \href{http://xxx.lanl.gov/abs/1010.1253}{{\tt arXiv:1010.1253}}.

\bibitem{Perez:2015ssa}
M.~G. Pérez et al.
{\em JHEP}
  {\bf 06} (2015) 193, [\href{http://xxx.lanl.gov/abs/1505.0578}{{\tt
  arXiv:1505.0578}}].

\bibitem{Gonzalez-Arroyo:2014dua}
A.~Gonzalez-Arroyo and M.~Okawa, 
{\em JHEP} {\bf 1412} (2014) 106,
  [\href{http://xxx.lanl.gov/abs/1410.6405}{{\tt arXiv:1410.6405}}].

\bibitem{Gonzalez-Arroyo:2015bya}
A.~Gonz\'alez-Arroyo and M.~Okawa, 
\href{http://xxx.lanl.gov/abs/1510.0542}{{\tt arXiv:1510.0542}}.

\bibitem{opencoarray:2014}
A.~Fanfarillo et al.
{\em Proceedings of the 8th International Conference on
  PGAS Programming Models. ACM} (2014).

\end{thebibliography}\endgroup

\end{document}